\newif\ifdraft\drafttrue
\documentclass[sigplan,twocolumn]{acmart}
\renewcommand\footnotetextcopyrightpermission[1]{}
\acmConference[arXiv Preprint]{arXiv preprint}{October 2025}{arxiv.org}

\usepackage{tikz}
\usepackage{amsmath}

\usepackage{xspace}
\usepackage{pgfplots}
\usepackage{cleveref}
\usepackage[frozencache]{minted2}
\usepackage{subfigure}
\usepackage{simplebnf}
\usepackage{booktabs}
\usepackage{enumitem}
\usepackage{adjustbox}
\usepackage{dsfont}
\usepackage{wrapfig}
\usepackage{algpseudocode}
\usepackage{algorithm}
\usepackage[most]{tcolorbox}
\usepackage{xcolor}

\usetikzlibrary{arrows.meta, positioning, patterns}
\usepgflibrary{shapes.symbols}
\usetikzlibrary{shapes.symbols} 
\usetikzlibrary{decorations.pathreplacing, shapes}
\input{misc}

\newcommand{\sys}{\textsc{Caruca}\xspace}

\newcommand{\heading}[1]{\vspace{2pt}\noindent\textbf{\emph{#1}}:\enspace}

\newcommand{\ttt}[1]{\texttt{#1}}
\newcommand{\unix}{{\scshape Unix}\xspace}
\newcommand{\chlng}[1]{\textnormal{#1}\xspace}

\ifdraft

\newcommand{\kk}[1]{[{\color{red}kk: #1}]}
\newcommand{\sh}[1]{[{\color{orange}sh: #1}]}

\else

\newcommand{\kk}[1]{}
\newcommand{\sh}[1]{}

\fi

\def\eg{{\em e.g.}, }
\def\ie{{\em i.e.}, }
\def\etc{{\em etc.}\xspace}

\crefformat{figure}{Fig.~#2#1#3}
\crefformat{table}{Tab.~#2#1#3}
\crefformat{section}{\S#2#1#3}
\crefmultiformat{section}{\S#2#1#3}{--\S#2#1#3}{, \S#2#1#3}{ and \S#2#1#3}

\pgfplotsset{compat=1.18}

\crefformat{section}{\S#2#1#3}
\crefrangeformat{section}{\S#3#1#4--\S#5#2#6}

\definecolor{clrinput}{RGB}{255,204,204}
\definecolor{clroutput}{RGB}{204,238,255}
\definecolor{clrcomponent}{RGB}{238,238,187}

\begin{document}

\date{}

\title[Caruca: Effective and Efficient Specification Mining for Opaque Software Components]{Caruca: Effective and Efficient Specification Mining\\ for Opaque Software Components}

\author{Evangelos Lamprou}
\email{vagos@lamprou.xyz}
\affiliation{
  \institution{Brown University}
}

\author{Seong-Heon Jung}
\email{sj4963@nyu.edu}
\affiliation{
  \institution{New York University}
}

\author{Mayank Keoliya}
\email{mkeoliya@upenn.edu}
\affiliation{
  \institution{University of Pennsylvania}
}

\author{Lukas Lazarek}
\email{lukas\_lazarek@brown.edu}
\affiliation{
  \institution{Brown University}
}

\author{Konstantinos Kallas}
\email{kkallas@ucla.edu}
\affiliation{
  \institution{UCLA}
}

\author{Michael Greenberg}
\email{michael@greenberg.science}
\affiliation{
  \institution{Stevens Institute of Technology}
}

\author{Nikos Vasilakis}
\email{nikos@vasilak.is}
\affiliation{
  \institution{Brown University}
}

\renewcommand{\shortauthors}{E. Lamprou, S. Jung, M. Keoliya, L. Lazarek, K. Kallas, M. Greenberg, N. Vasilakis}

%
%
%
%
%
%

%
%
%

\begin{abstract}
%
%
%
%
A wealth of state-of-the-art systems demonstrate impressive improvements in performance, security, and reliability on programs composed of opaque components, such as \unix shell commands.
To reason about commands, these systems require partial specifications.
However, creating such specifications is a manual, laborious, and error-prone process, limiting the practicality of these systems.
This paper presents \sys, a system for automatic specification mining for opaque commands. 
To overcome the challenge of language diversity across commands, \sys first instruments a large language model to translate a command's user-facing documentation into a structured invocation syntax.
Using this representation, \sys explores the space of syntactically valid command invocations and execution environments.
\sys concretely executes each command-environment pair, interposing at the system-call and filesystem level to extract key command properties such as parallelizability and filesystem pre- and post-conditions.
These properties can be exported in multiple specification formats and are immediately usable by existing systems.
Applying \sys across 60 GNU Coreutils, POSIX, and third-party commands across several specification-dependent systems shows that \sys generates correct specifications for all but one case---completely eliminating manual effort from the process and currently powering the full specifications for a state-of-the-art static analysis tool.
\end{abstract}

\maketitle

\begin{figure}[t]
    \centering
    \begin{adjustbox}{max width=.9\columnwidth}
    \begin{tikzpicture}[
        box/.style={rectangle, draw, solid, minimum width=2cm},
        multi/.style={draw, rectangle, densely dashed, inner sep=0.1cm},
        arrow/.style={-{Latex[length=3mm, width=2mm]}, thick},
        dashedline/.style={densely dashed},  
        circle-label/.style={fill=white, draw, circle, inner sep=0.5mm},
        subbox/.style={rectangle, draw, solid, minimum width=2cm}
    ]
        \node[box, minimum height =.7cm, rounded corners, fill=clrinput] (binary) {Command Binary};
        \node[doc, draw, below=.5cm of binary, align=center, solid, fill=clrinput] (documentation) {Documentation};
        \node[multi, below=1cm of documentation] (syntax-group) {
            \begin{tikzpicture}[
              box/.style={rectangle, draw, solid, minimum width=4cm}
              ]
                \node[box, minimum width=1cm, minimum height =.7cm, rounded corners, fill=clrcomponent] (parser) {LLM Parser};
                \node[tape, draw, below=0.5cm of parser, minimum height =.7cm, solid,fill=clroutput] (stxspec) {Syntax Spec};
                \draw[arrow, solid] (parser) -- (stxspec);
            \end{tikzpicture}
        };
        \node[left=0.05cm of syntax-group, align=center] (syntax-label) {Syntax \\ Inference};
        \node[circle-label] at (syntax-group.north east) {\cref{sec:syntax-spec}};
        
        \node[box, below=0.5cm of syntax-group, minimum height=.7cm, minimum width=2.75cm, align=center, fill=clrcomponent] (generator) {Configuration \\ Generator};

        \node[circle-label] at (generator.north east) {\cref{sec:config-gen}};
        
       \node[box, below=0.5cm of generator, minimum width=3cm, fill=clrcomponent] (generated) {
    \begin{tikzpicture}
        \node[subbox, fill=white] (cmd) at (0,0.6) {Command Invocations};
        \node[subbox, fill=white] (fs) at (0,-0.3) {Execution Environments};
        \node at (0,0.15) {\( \times \)};
    \end{tikzpicture}
};

        \node[multi, right=1.7cm of binary, xshift=0.75cm, yshift=-0.4cm, fill=clrcomponent] (tracer-group) {
            \begin{tikzpicture}
                \node[box, fill=white] (tracer1) {Syscalls};
                \node[box, below=0.2cm of tracer1, fill=white] (tracer2) {FS {Diff-ing}};
            \end{tikzpicture}
        };
        \node[above=0.1cm of tracer-group] {Tracer};
        
        \node[circle-label] at (tracer-group.north east) {\cref{sec:tracing}};
        
        \node[multi, below=1.1cm of tracer-group, fill=clrcomponent] (annotator-group) {
            \begin{tikzpicture}[
            box/.style={rectangle, draw, solid, minimum width=4cm}
            ]
                \node[box, fill=white] (annotator1) {File I/Os \cite{pash_osdi, posh2020, shash}};
                \node[box, below=0.2cm of annotator1, fill=white] (annotator2) {Parallelizability \cite{pash_osdi, posh2020}};
                \node[box, below=0.15cm of annotator2, fill=white] (annotator3) {I/O Size Relation \cite{posh2020}};
                \node[box, below=0.15cm of annotator3, fill=white] (annotator4) {FS Side-effects \cite{shash, holen_shellcheck_2024}};
            \end{tikzpicture}
        };
        \node[above=0.1cm of annotator-group, xshift=1.2cm, align=left] {Specification \\ Derivation};

        \node[circle-label] at (annotator-group.north east) {\cref{sec:specs}};
        \node[tape, draw, below=0.5cm of annotator-group, align=center, fill=clroutput] (annotation) {Command Specs};
        \node[box, draw, below=0.5cm of annotation, rounded corners, fill=clrcomponent] (adapter) {Adapters};

        \node[box, draw, below=0.5cm of adapter] (consumers) {Consumers \cite{pash_osdi, posh2020, shash, holen_shellcheck_2024}};
        \node[circle-label] at (consumers.north east) {\cref{sec:eval}};

        \draw[arrow] (binary.east) to[out=0, in=180] (tracer-group.west);       
        \draw[arrow] (documentation) -- (syntax-group);
        \draw[arrow] ([yshift=0.2cm]syntax-group.south) -- (generator);

        \draw[arrow] (annotation) -- (adapter);

        \draw[arrow, dotted] (adapter) -- (consumers);

        \draw[dashedline] (generator.south west) -- (generated.north west);
        \draw[dashedline] (generator.south east) -- (generated.north east);
        \draw[arrow] (generator.east) to[out=0, in=180] (tracer-group.west);       
        \draw[arrow] (tracer-group) -- (annotator-group);
        \draw[arrow] (annotator-group) -- (annotation);
    \end{tikzpicture}
    \end{adjustbox}
    \caption{\textbf{\sys architecture overview.}
\sys takes as input a command binary and its documentation, and first creates a syntax specification (\cref{sec:syntax-spec}).
From this, its configuration generator (\cref{sec:config-gen}) creates diverse invocations and environments.
These are executed in a traced sandbox (\cref{sec:tracing}), recording all system-level interactions.
Finally, the specification derivation subsystem (\cref{sec:specs}) analyzes the collected traces to produce specifications for downstream systems (\cref{sec:eval}).
Colors indicate component roles: \colorbox{clrinput}{red}---static inputs; \colorbox{clrcomponent}{green}---subsystems; \colorbox{clroutput}{blue}---runtime and analysis outputs.
    }
    \label{fig:sys-arch}
\end{figure}
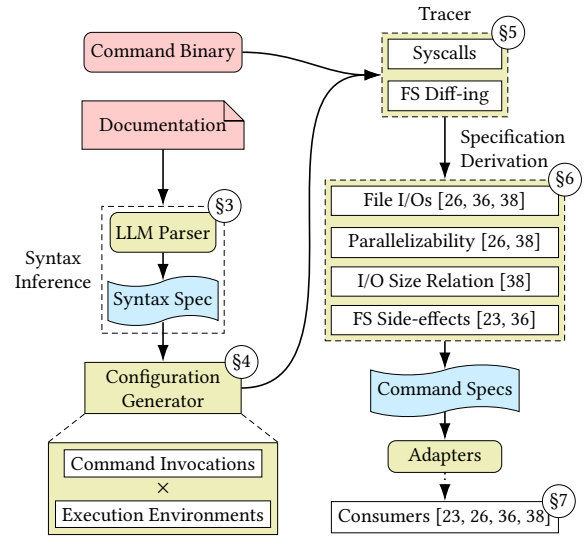

\section{Introduction}
\label{intro}
Command-line utilities (or commands in short) are an integral part of \unix~\cite{unixenv} and other environments~\cite{nosilverbullet,lampson2004software}.
The key characteristic of commands is that they allow for abstraction and composition---they are written in a variety of programming languages, are often distributed as opaque binaries, and enclose versatile, well-tested, and extensively documented functionality that can be easily interfaced with.
Take as an example the script shown in \Cref{fig:example-script}---the script's entire functionality involves the composition of four commands: \ttt{rm}, \ttt{cp}, \ttt{uglifyjs}, and \ttt{gzip}.

Due to the popularity and importance of such scripts, there has been a wealth of state-of-the-art systems focusing on improving their performance~\cite{pash_osdi,posh2020,dish}, security~\cite{abash}, and reliability~\cite{shash,becker:hal-03737886}.
These systems must reason about both a script's execution and the commands it composes.
However, these commands could be written in arbitrary languages and their source code might not be available, making it infeasible to use any type of static source-code analysis.
So far, the solution to this has been handwritten annotations---partial specifications that are painstakingly written by the system authors to capture important aspects of command behavior.
Examples of such specifications include parallelizability properties~\cite{pash_osdi,posh2020}, monotonicity semantics~\cite{posh2020}, and Hoare-style filesystem pre- and post-conditions~\cite{shash,becker:hal-03737886}.
Transferring these benefits beyond these systems' limited evaluation sets requires scaling this specification effort---these papers suggest crowd-sourcing \emph{hundreds of thousands} of command instances---a number further exacerbated by differences across command versions on Linux, macOS, BSD, \etc, as well as initiatives to rewrite classic utilities in safer languages~\cite{ubuntuoxidising}.
This effort faces severe human-effort, scalability, and correctness concerns---and hence has not materialized.

To address this issue, this paper presents \sys, a system for generating specifications for opaque software components, focusing on shell commands.
\sys supports components written in any language, making no assumptions or requirements about source code availability, and produces a variety of different specifications---targeting performance optimizations, correctness checks, and other systems.
\sys leverages the extensive documentation provided with these components and their observable interactions they have with the broader environment as the only common denominators.

\sys begins (\cref{fig:sys-arch}) with its invocation syntax inference engine, which infers a structured specification of a command's invocation syntax from its natural language documentation---\eg \ttt{man} pages, markdown files, and other sources of documentation.
Using the syntax specification, \sys's configuration generator creates a large number of test invocations and environments, sweeping through the possible flags, options, arguments, and filesystem states.
\sys's tracer then instantiates concrete environments and executes each command configuration with appropriate interposition recording all of the command's interactions within its environment.
Finally, \sys's command specification derivation subsystem examines the traces extracted by the concrete executions and applies a series of transformation rules to produce \emph{command specifications}.
All but the last of these components are common across all different command specification types and downstream systems.

\sys is evaluated---with particular emphasis on its components that depend on LLMs---on 60 GNU Coreutils, POSIX, and third-party commands for which ground-truth specifications exist in prior work~\cite{posh2020, pash2021eurosys, shash, holen_shellcheck_2024} and an extended set of 120 commands with manually derived syntax specifications.
\sys successfully generates correct specifications for 59/60 commands, discovering additional constraints missed by the original handwritten specifications.
\sys produces partial specifications within 1 hour for 103 out of 120 commands and within 24 hours for all but one command.
These results place \sys as the first fully automated and generally applicable specification miner for opaque components.
\sys already powers the full specifications for Shseer~\cite{sash-hotos25}, a state-of-the-art static analysis tool~\cite{shash}.

\heading{Outline and contributions}
The paper starts with an example outlining key challenges in automated specification inference~(\cref{sec:example}).
It then proceeds with \sys's contributions:
\begin{itemize}[leftmargin=*, topsep=0pt, itemsep=2pt, partopsep=0pt, parsep=0pt]

\item \textbf{Invocation syntax specification and inference}~(\cref{sec:syntax-spec}): \\
\sys encodes an inferred model of a command's interface in a \emph{syntax specification}, using a domain specific language (DSL) designed to eliminate ambiguity and enable effective configuration generation.

\item \textbf{Environment model and generation}~(\cref{sec:config-gen}):
\sys models key aspects of the filesystem and uses this model to generate diverse command invocations that combine explicit command arguments and implicit system state.

\item \textbf{Tracing}~(\cref{sec:tracing}) \textbf{and specification derivation}~(\cref{sec:specs}): \sys executes commands with system-level instrumentation and introduces rules for summarizing and translating the extracted traces to command specifications.

\end{itemize}

\noindent
The paper then presents an evaluation of \sys~(\cref{sec:eval}) and concludes with a discussion of related work (\cref{sec:related-work}).

\heading{Availability}
\sys will be available as an MIT-licensed open source artifact available for download at: 
\begin{center}
    \url{https://github.com/binpash/caruca}
\end{center}

\begin{figure}[t]
    \begin{minted}[fontsize=\small]{bash}
# Clean the destination directory
rm -rf dist/*                  
# Copy source file to dist
cp src/htmx.js dist/htmx.js       
# Minify script
uglifyjs -m eval -o dist/htmx.min.js dist/htmx.js        
# Compress minified script
gzip -9 -k -f dist/htmx.min.js > dist/htmx.min.js.gz            
    \end{minted}
    \caption{\textbf{Example shell script from the \ttt{htmx} project~\cite{htmx}.}
    The script prepares the \ttt{htmx} library for download. It copies \ttt{htmx}'s source code inside a directory, minifies it, and compresses it.
    }
\label{fig:example-script}
\end{figure}

\section{Overview}
\label{sec:example}

\Cref{fig:example-script} shows a script excerpt from the \ttt{htmx} front-end library~\cite{htmx}.
The script prepares the library's source code for download:
  it first cleans the destination directory,
  copies inside the program's source code using \ttt{cp},
  minifies the program using \ttt{uglifyjs}, and finally 
  compresses the minified program using \ttt{gzip}.
As this script directly affects the performance and reliability of client-facing JavaScript code, organizations that use \ttt{htmx} may seek to reduce the script's execution time and ensure its reliability.
Systems such as PaSh~\cite{pash_osdi}, POSH~\cite{posh2020}, Shellcheck~\cite{holen_shellcheck_2024}, and Shseer~\cite{sash-hotos25} achieve this goal:
  PaSh can parallelize the script on a multicore computer,
  POSH can scale it out across multiple computers,
  Shellcheck can identify dubious code patterns,
  and Shseer exposes subtle filesystem bugs.
To analyze arbitrary opaque commands and utilities that could be written in different languages, these systems rely on the use of manually written command specifications that describe aspects of their behavior, \eg their parallelizability, system dependencies, and filesystem effects.
Unfortunately, all these systems currently lack annotations for at least one of the commands in this script (\ttt{uglifyjs}, \ttt{gzip}, \ttt{rm}, and \ttt{cp}) and therefore are not able to analyze it. 
Right now, a user would have to manually write the missing specifications.
This missing piece limits the practical applications of all these systems.

\sys addresses this problem and frees developers from having to manually write command specifications by automating this process.

\subsection{Challenges}
\label{sec:challenges}


Automatically inferring command specifications, however, involves several challenges that \sys needs to address:

\heading{\chlng{(C1)} Commands lack structured, typed interfaces}
In order to adequately explore a command's execution space, \sys needs to generate a wide range of arguments to invoke the command with.
However, in contrast to typed languages, where type information constrains the space of valid inputs (\eg \ttt{5} is an \ttt{int} and \ttt{"foo"} is not), the interface of commands is untyped: arguments are represented as an array of strings. 
Each command dynamically parses and checks the validity of its arguments, with no standard way for a command to inform that its arguments were invalid, \eg via an exit status or an error message.
This makes it extremely challenging to generate valid invocations for commands: using arbitrary strings is extremely unlikely to lead to valid executions and it is generally impossible to distinguish between executions that return an error exit code due to failed argument parsing or other reasons.

\heading{\chlng{(C2)} Command behavior combinatorial explosion}
In order to generate a complete specification of a command's behavior, \sys needs to explore all possible execution modes of a command by invoking it with different combinations of flags and options.
Even a simple command invocation like \ttt{rm -rf} demonstrates how flags and options lead to a combinatorial explosion of command behaviors.
The basic \ttt{rm path} invocation deletes a file at the specified path but fails if the path is not a file or does not exist. 
Using the \ttt{-r} flag, \ttt{rm -r path} deletes directories and their contents recursively, failing only if the path does not exist. 
With the \ttt{-f} flag, \ttt{rm -f path} forces deletion, ignoring errors such as an nonexistent path.
Combining both flags, \ttt{rm -rf path} performs a forceful, recursive deletion of the specified path.
Therefore, each possible flag combination demands a new corresponding specification.
This variation means that a command has potentially an exponential number of behaviors relative to the number of flags, making the problem of specification inference through command execution intractable.

\heading{\chlng{(C3)} Implicit command dependencies}
In addition to flags, options, and arguments, a command's behavior also \emph{implicitly} depends on the state of the filesystem.
For instance, \ttt{rm path} changes behavior depending on the state of \ttt{path}:
if it does not exist or is a directory, \ttt{rm} takes no action and exits with code \ttt{1};
if the path exists and is a file it removes it and exits with code \ttt{0}.
%
To explore a command’s behavior, \sys must identify relevant filesystem states and generate corresponding invocations to run against them.

\heading{\chlng{(C4)} Command execution isolation and monitoring}
Finally, \sys needs to execute and monitor a large set of command invocations in order to be able to later infer their specifications.
These invocations need to run in a custom environment so that \sys can test their behavior, but at the same time they must be isolated so that their execution does not affect the surrounding system or other---under exploration---invocations.
Not only that, but since the number of invocations is very large, this isolation and monitoring needs to add minimal overhead on the command execution.
%

\subsection{System Overview}

\sys tackles these challenges through four components: 
(C1) a syntax specification language and inference engine that extracts argument structure from documentation; 
(C2) a configuration generator that heavily prunes the invocation space;
(C3) a filesystem model that pairs invocations with environments to form configurations; and
(C4) a tracer that interposes on command execution using lightweight sandboxing and tracing.
This section illustrates these components with the \ttt{rm} command.

\heading{Invocation Syntax Inference (\cref{sec:syntax-spec})}
\sys starts by generating a syntax specification for the command, which describes all of its flags, options, and positional arguments, as well as the types of all arguments (option and positional).
In the case of \ttt{rm}, \ttt{-r} and \ttt{-f} are marked as \ttt{Flag}s, while the positional arguments that will be used in the invocation after globbing, \ie \ttt{*} resolution, are assigned the \ttt{Path} type.
The specification is expressed as a language embedded in Python and describes the syntactically correct ways to invoke the command.
\sys generates the specification by using \ttt{rm}'s documentation, \ie its \ttt{man} page or \ttt{-{}-help} output, and passing it to a large language model prompted to generate syntax specifications.
The LLM solely generates the syntax specification---it is not involved in any remaining components that observe command execution and derive command specifications.
Given as input \ttt{rm}'s \ttt{man} page, \sys arrives at the following syntax specification (truncated).
\begin{minted}[fontsize=\small,xleftmargin=\parindent]{python}
rm_s=[[Flag("-r"), Flag("-f")],[Path(arity='1+')]]
\end{minted}

\heading{Configuration Generator (\cref{sec:config-gen})}
Guided by the type and argument information from the syntax specification,
\sys's generator produces command invocations that are syntactically valid and well-typed.
A study on a very large set of shell scripts from GitHub determines that 99.998\% of all the invocations found in this set only have up to 4 flags and options~(\cref{sec:cmd-invocations}).
\sys uses this insight to prune the invocation space to up to 4 flags and options, but can be configured to either (1) generate invocations with more flags, or (2) generate specific invocations to infer targeted specifications for them.
In addition, \sys pairs each invocation string with a set of diverse filesystem states to adequately explore implicit filesystem dependencies for command behavior.
The invocation string and the system state jointly form an invocation configuration.
For \ttt{rm}, \sys generates 8,749,056 invocation configurations, taking into consideration type-specific variations such as alternative argument strings, path arguments pointing to a file, directory, nothing, \etc

\heading{Tracing (\cref{sec:tracing})}
\sys then executes and traces each invocation configuration.
The executions take place inside a custom, lightweight sandbox.
Traces include a command's system calls and its modifications to the filesystem.
The simplified trace corresponding to the \ttt{rm path} invocation under an execution environment where \ttt{path} points to a regular file is \ttt{\{read path, delete path\}}.

\heading{Command Specification Derivation (\cref{sec:downstream})}
The collected traces contain all of the information about the command's interactions with the surrounding system and can therefore completely describe the behavior of each tested invocation.
\sys collects these traces and performs a lightweight analysis on them to infer a complete command specification, capturing properties that are important for all downstream systems, including the command's inputs, outputs, filesystem effects, and parallelizability. 
Concretely, \cref{fig:truncated-rm-annotation} illustrates the specification for \ttt{rm} generated by \sys (truncated for presentation).
This universal specification is then exported to downstream systems, enabling them to analyze, optimize, and find bugs in the target script.



\begin{figure}[t]
  \centering
\begin{align*}
  \mathrm{flags}(\{\}) \land \mathrm{arg}(\ttt{\$1}, \ttt{p.FD})
  \implies\\
  \nexists\, \ttt{\$1} \land \mathrm{exit}(0)
  \land I(\{\ttt{\$1}\})
  \land O(\{\}) 
  \land \mathcal{P}(\ttt{SE})
  \land  \not \downarrow
  \\
  \ldots
  \\[.5em]
  \mathrm{flags}(\{-r, -f\}) \land
  \mathrm{arg}(\ttt{\$1}, \ttt{p.FD} \lor \ttt{p.DIR} \lor \ttt{p.}\epsilon)
  \implies \\
  \nexists\, \ttt{\$1} \land \mathrm{exit}(0)
  \land I(\{\ttt{\$1}\})
  \land O(\{\})
  \land \mathcal{P}(\ttt{SE})
  \land \not \downarrow
  \\
  \ldots
\end{align*}
\vspace{-2em}
    \caption{\textbf{Specification for \ttt{rm}.} Excerpt from the \ttt{rm} specification produced by \sys.
    The specification encodes per-invocation pre- and post-conditions related to the command's filesystem effects, the way it interacts with its input ($I$) and output ($O$) streams, alongside summaries of its parallelizability ($\mathcal{P}$) and monotonicity ($\downarrow$) properties~(\cref{sec:specs}).}
    \label{fig:truncated-rm-annotation}
\vspace{-1em}
\end{figure}

\heading{Usage}
\sys is invoked to generate the required specifications for each command (given that they are executable and available in the \ttt{PATH}) as follows:
\begin{minted}[fontsize=\small]{bash}
$ man rm | caruca > rm.spec
$ man cp | caruca > cp.spec
$ uglifyjs --help | caruca > uglifyjs.spec
$ man gzip | caruca > gzip.spec
\end{minted}
\noindent
\sys has two expected application scenarios.
\sys can be invoked once per command after its creation, producing a specification that can be versioned and subsequently reused across target scripts and downstream systems
\sys can also be invoked in a targeted mode, generating a specification for a specific invocation, with some of the flags, options, arguments, and environment given concrete values.
This could be useful if a user just wants to do a one-off analysis or optimization of a single script with a single instance of a command that they are not going to use in other setups.

\heading{Limitations}
\sys currently has the following limitations and threats to validity.
First, even though it carefully generates relevant command invocations, through its command usage study, filesystem model, and argument types, it still only generates a \emph{finite} number of invocation configurations, which means that there could be cases, \eg if a command is invoked with more flags than what \sys explores, that a specification is incomplete.
To assess this probability, the evaluation includes a comprehensive study of the coverage of \sys's generated invocations on real-world scripts (\cref{sec:eval:completeness}).
Furthermore, as described above, users can use \sys in a targeted mode to generate a specification for a specific invocation, \eg using all the command flags, if they expect those invocations to be important.
Second, the syntax specification generator uses a statistical component (LLM) and therefore even though the task is simple, \ie determine type information about the flags, options, and arguments from a command's documentation, it does not come with formal correctness guarantees.
We assess the correctness of \sys's syntax generator for 120 commands from various suites (\cref{sec:eval:llm-parser}) and show that it is correct across 99.7\% of all flags, options, and arguments, with mistakes being non-catastrophic.
Finally, \sys does not generate environment variables in its invocation configurations, so it is agnostic to command behavior changes based on the values of environment variables.

\section{Invocation Syntax Inference}
\label{sec:syntax-spec}


Commands lack a well-typed invocation interface (C1, \cref{sec:challenges}); without knowing valid flags, options, argument types, or arity, exploring their invocation space reduces to generating arbitrary strings that mostly yield errors.
To address this, \sys defines a DSL that captures a command's syntax---flags, options, positional arguments, and type constraints---which serves as an effective grammar for invocation generation.
Since source code is unavailable, \sys instead leverages natural language documentation (\eg \ttt{man} or \ttt{-{}-help} pages) and iteratively prompts an LLM to extract the corresponding syntax specification.

\subsection{Syntax Specification DSL Design}

\sys's syntax specification DSL captures the following information about a command:
(1) its flags,
(2) its options and the types of their option arguments,
(3) its positional arguments together with their type and arity.

\heading{Grammar}
\sys tackles this by storing information about the valid ways of using commands as a DSL.
At the top level, the \textit{Command} element contains the name of the command and a set of possible grammars for using the command, represented by \textit{Usage}.
A \textit{Usage} is an ordered sequence of \textit{Position}s, which represent the positions where arguments can be located.
A \textit{Position} contains a set of \textit{Argument}s which can appear in that position.
Note that the ordering of \textit{Position}s is significant, while the ordering of \textit{Arg}s in a single \textit{Position} is not.
For a command like \ttt{mv}, the first \textit{Position} holds flags such as \ttt{-v} and \ttt{-f}, while the second and third \textit{Position}s hold the source and destination paths, which are treated differently (\eg \ttt{mv -v -f src dst}).
The \textit{Arg} element contains an \textit{Arity} value and a \textit{Type}.
The \textit{Arity} of an argument can be \ttt{zero\_one}, \ttt{zero\_plus}, \ttt{one\_plus}.
The \ttt{zero\_one} corresponds to an optional argument, while \ttt{zero\_plus} and \ttt{one\_plus} represent arguments that can appear at least zero times or at least one time respectively.
Alternatively, if an argument must appear a fixed number of times, it will have arity $n$, where $n$ is some concrete positive natural number.
In addition to these core elements, the \textit{Arg} element accept the options \ttt{flag\_followed\_by\_equals} (the command expects the option's values to be preceded by an equals sign), \ttt{dash\_as\_stdin} (the command interprets a dash as \ttt{/dev/din}), and \ttt{max\_repetition} (the command accepts multiple instances of this flag/option), which reflect nuances in how the command parses that flag/argument.
Each argument also takes a set of aliases, which are alternative strings that can be used to refer to the same argument.

\heading{Argument types}
Lastly, \textit{Type} expresses the valid formats for the argument.
While all arguments can be represented as string, in practice a lot of them are not actually arbitrary but can be constrained in various ways, \eg only representing an integer, path, or values from a range.
To arrive at a set of ubiquitous argument types useful across a variety of commands, two domain experts studied the documentation of 120 command from various sets, including GNU coreutils and commands currently supported by PaSh, POSH, Shellcheck, and Shseer.
This process took over 200 person-hours.
The resulting set of types are
\ttt{path}, which represents a string that is a valid POSIX path, absolute or relative;
\ttt{selection}, an enumeration type where the valid values are a finite set of strings;
\ttt{integer} and \ttt{char}, which represent integers and single characters respectively,
\ttt{string}, an arbitrary string,
and \ttt{other} as a fallback type for when \sys's LLM component is not able to infer a more suitable type.
If not specified otherwise, arguments of type \ttt{other} are assumed to be of type \ttt{string}.
\sys's library is easily extensible to support more types.
This paper leaves an extensive characterization of all argument types for future work.

\subsection{Documentation to Syntax Specification}
\label{sec:llm-parser}

To generate command syntax specifications, \sys relies on command documentation (\ttt{man} or \ttt{-{}-help} pages).
These specifications capture only argument structure, which is typically unambiguous, unlike usage examples or behavioral descriptions.
\sys prompts an LLM---prompted to be positioned as a syntax expert and constrained to the DSL's argument types---with three examples (\ttt{rm}, \ttt{mv}, \ttt{touch}) alongside the available documentation, to produce the specification.




\noindent
After the LLM generates the syntax specification, \sys loads the generated specification as a Python module and validates it for syntactic and type correctness.
If the specification is invalid, \sys retries the transformation, provides the previous output as context and includes the error message in the prompt.
After three failed attempts, \sys will issue an error message and abort the transformation.
The number of retries is configurable, but in practice three attempts have been sufficient for all commands that \sys has been applied to.
\sys uses the GPT-4o model for this task, but the process itself is model-agnostic.

\section{Configuration Generation}
\label{sec:config-gen}

Given a command syntax specification, \sys generates an extensive set of command invocations with varying arguments and environments to explore the command behaviors.
These invocations are executed and traced in an isolated environment to determine the behavior and specification of a command.
This section describes how \sys generates the set of invocations and environments for each command and how it prunes that space to keep generation and later execution tractable.


\subsection{Flag and option study}
\label{sec:flag-study}
\begin{figure}[t]
  \includegraphics[width=\columnwidth]{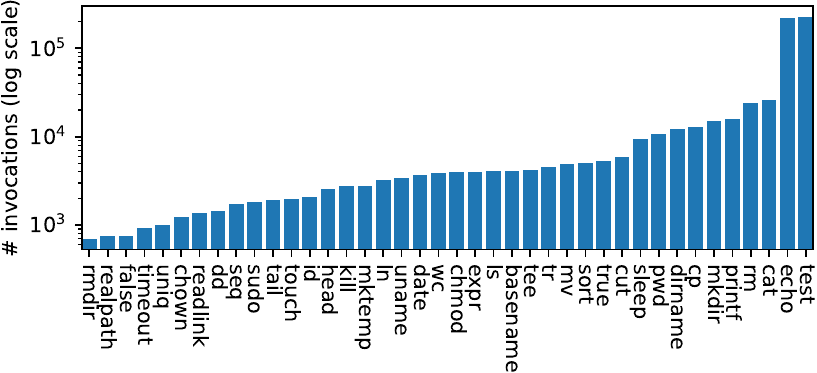}
  \caption{\textbf{Histogram of command invocation counts.} Histogram of the number of invocations per command in the collected dataset for the top 40 most invoked commands.
  }
  \label{fig:invdistr}
  \vspace{-1em}
\end{figure}

\sys's pruning strategies are informed by a study on real-world shell scripts collected from GitHub.
These script come from repositories with more than 10 stars that use the shell as their primary language.
The study includes over 49K shell scripts, totaling 19 million lines of code.
Each of the scripts is parsed and all its command invocations are extracted, totaling over 665K invocations.
For each invocation, its number of flags and options is counted.
%
The results of the study are that out of all the invocations there are:
430,680 (64.7\%) invocations with no flags or options,
193,380 (29.0\%) invocations with 1 flag or option,
39,616 (5.9\%) invocations with 2 flags or options,
1,647 (0.2\%) invocations with 3 flags or options,
282 (0.04\%) invocations with 4 flags or options, and
9 (0.00135\%) invocations with 5+ flags or options.
These results mean that by just exploring invocations with at most 2 flags, \sys covers 99.6\% of the total invocations.
Therefore, it can prune invocations with more flags if their exploration is intractable.
By default \sys only explores a command's invocations with up to 4 flags and options, but this number is configurable.
\Cref{fig:invdistr} shows the distribution of the number of invocations per command in the invocation dataset for the top 40 most invoked commands.

\subsection{Command Invocations}
\label{sec:cmd-invocations}

\sys generates a wide variety of command invocations to explore each command's behaviors being guided by the syntax specification.
\sys creates invocations exploring (1) combinations of flags and options and (2) several possible values for each positional and option argument.


A key challenge that \sys needs to address is that there is an exponential number of possible invocations for a command (C2, \cref{sec:challenges}) and this could make generation and later execution intractable.
\sys addresses this challenge by constraining the set of inputs in two ways: (1) it uses the argument type information provided in the syntax specification to only generate well-typed argument instances, and (2) it limits the number of flags and options in each invocation based on a study of real-world command usage.

\heading{Typed argument generators}
\sys balances exploration completeness and tractability by using the type information described in the syntax specification DSL~(\cref{sec:syntax-spec}). Each type is accompanied by an input generator that generates a sample of valid and well-typed values for that type.
For example, for the \ttt{selection} type, \sys generates a value for all its possible values.
For \ttt{integer}, it generates the numbers -1, 0, 1.
For \ttt{path}, it generates any of the possible path types included in \sys's filesystem model~(\cref{sec:exec-environments}).

\subsection{Execution Environments}
\label{sec:exec-environments}


Another challenge is that command behavior depends not only on arguments but also on filesystem state (C3).
For instance, \ttt{cp src target} succeeds only if \ttt{src} exists, requiring executions under both conditions.
For this, \sys pairs each invocation with environments generated from a filesystem model; these \emph{invocation configurations} specify both filesystem topology and file/stream contents.


\begin{figure}[t]
    \small
    \begin{bnf}
        \textit{Path} ::= (\textit{PathType}, \textit{PathPointer});;
        \textit{PathPointer} ::=
        | \ttt{fs} (FileSystem)
        | \ttt{nonexistent}
        | \ttt{parent\_nonexistent} ;;
        \textit{FileSystem} ::=
        | \ttt{directory} (List<\textit{FileSystem}>)
        | \ttt{file};;
        \textit{PathType} ::= 
        | \ttt{absolute}
        | \ttt{relative};;
    \end{bnf}
    \caption{\textbf{\sys's filesystem model.} 
    \sys's filesystem model describes the possible filesystem states that \sys considers relevant to command behavior.
    The model captures the filesystem at the path and object level.
    A filesystem object can be a file or a directory, and a filesystem path can either point to a filesystem object or be non-existent.
    }
    \label{fig:fs-model}
    \vspace{-1.5em}
\end{figure}

\heading{Filesystem topology}
The behavior of many commands depends on the state of the filesystem at the time of execution.
For example, commands like \ttt{mkdir} or \ttt{rm} behave differently depending on whether the target path exists, whether it is a file or directory, and whether its parent directory exists.
An invocation like \ttt{mkdir dir1/dir2} fails if the parent path \ttt{dir1} does not exist.
Providing the \ttt{-p} flag makes the command succeed in either case.
Similarly, \ttt{rm p} succeeds if \ttt{p} is a file, but fails if it is a directory unless the \ttt{-r} flag is provided.
Using the \ttt{-f} flag makes the command succeed even if the target path does not exist.
\sys defines and uses a rich filesystem model that can describe relevant filesystem configurations that trigger different command behaviors.




\sys generates filesystem states using the model in \cref{fig:fs-model}, which operates at the path-string and filesystem-object levels.
Generation is done by exploring all combinations of the model's components.
At the string level, paths may be absolute or relative, and at the filesystem level, either nonexistent or mapped to objects.
For nonexistent paths, \sys creates both a missing child and a missing grandchild (with a nonexistent parent) to capture cases like \ttt{mkdir}, which succeeds on the former but fails on the latter.
When creating existing paths, \sys generates environments where the path is a file, or a directory (empty or containing one child).
This model suffices for useful specifications~(\cref{sec:eval}), though it can be extended (\eg with symbolic links or pipes) to capture richer preconditions and effects.

\heading{File/Stream content}
When populating files or \ttt{stdin} with content,
\sys also applies variations to help explore a broader space of behaviors, since some commands' behavior is dependent on their input.
For instance, running \ttt{grep "pattern"} on an input without the string \ttt{pattern} fails to reveal \ttt{grep}'s primary purpose.
%
\sys provides four types of content: text, math formulas, JSON, and image data.
Each content is constructed automatically by \sys, given a corresponding oracle.
For text, it is the corpus of Project Gutenberg,
for math calculations, it is a set of simple arithmetic expressions generated from a grammar that includes the four basic arithmetic operations and parentheses,
for JSON it is a set of JSON objects with varying structures sourced from a public dataset~\cite{json-generator}, 
and for images it is a set of PNG images ranging from 2kb to 4MB from the public domain.
\sys also generates partial variants of the content.
%




\section{Isolated Tracing}
\label{sec:tracing}

After \sys generates pairs of command invocations and accompanying execution environments, it iterates over each invocation configuration, dynamically creates the state described by it, and executes the concrete command invocation under said environment.
For this, \sys uses two key techniques: filesystem sandboxing and system-call tracing.

\heading{Filesystem sandboxing}
First, \sys executes commands inside a lightweight isolation layer using OverlayFS~\cite{overlayfslinux} that prevents filesystem side-effects from leaking into the host, and between different command executions.
\sys mounts each top-level directory (\ttt{/*}) into the sandbox with a writable \emph{upperdir} and the original directory as a read-only \emph{lowerdir}.
Any modifications made by the command are redirected to the upperdir without affecting the original directory.
In addition, \sys mounts the \ttt{tty}, \ttt{null}, \ttt{zero}, \ttt{full}, \ttt{random}, and \ttt{urandom} devices and remounts \ttt{/proc} freshly.
\sys leverages process isolation by entering a fresh Linux namespace~\cite{linuxnamespaces}, with mount, user, and pid isolation enabled.
After entering the namespace, \sys enters the sandbox using \ttt{chroot}---changing the root directory to the sandbox's root---and finally executes the command.
After the command finishes, \sys determines its filesystem effects by comparing the sandbox's upperdir (containing only modifications) against the host's original filesystem.
Hence, only the upperdir needs to be traversed, which is typically small, making this an efficient operation.
During traversal, \sys classifies every difference based on the filesystem model~(\cref{sec:exec-environments}) as file created, file modified, file removed, file replaced with directory, directory created, directory removed, or directory replaced with file.
Finally, \sys deletes the sandbox environment before moving to the next invocation configuration.

\heading{System-call tracing}
To observe more fine-grained and ordered command behavior, useful for inferring certain command properties, \sys attaches a system-call tracer to the command, recording all system calls made during its execution using \ttt{strace}~\cite{strace}.
To reduce tracing overhead, \sys traces only a narrow set of system calls relevant to filesystem activity and location probing, avoiding the cost of recording all calls.
In addition, it uses seccomp-BPF~\cite{seccomp} to filter out irrelevant system calls in-kernel, preventing them from crossing into user space and being recorded by \ttt{strace}. 
After tracing is complete, \sys converts the system-call traces into a structured format, recording the system-call name, its arguments, and its return value, alongside assigning a classification for each system call as \emph{read}~(\eg \ttt{open} with \ttt{O\_RDONLY}, \ttt{stat}, \ttt{statfs}, \etc) or \emph{write}~(\eg \ttt{open} with \ttt{O\_CREAT}, \ttt{write}, \ttt{unlink}, \etc).
This translation makes \sys's specification derivation subsystem~(\cref{sec:specs}) agnostic to the specific tracing technique, allowing traces from other tools.

%

The system-calls \sys traces depend on the specifications it targets~(\cref{sec:specs}), making tracing configurable.
Other specification types can be supported by extending the traced call set.

\section{Command Specification Derivation}
\label{sec:downstream}
\label{sec:specs}


From execution traces of all invocation configurations, \sys generates behavior specifications for four downstream systems: performance optimizers PaSh~\cite{pash_osdi} and POSH~\cite{posh2020}, and bug-finders Shellcheck~\cite{holen_shellcheck_2024} and Shseer~\cite{sash-hotos25}.
This section details their specification needs and how \sys derives them from traces.



\subsection{Performance Optimization Systems}

We first describe the performance optimization systems, describing the semantics of their specifications and how they are used by the systems for optimization.

\heading{PaSh}
PaSh~\cite{pash_osdi} is a parallelizing script-to-script compiler.
It relies on command specifications\footnote{The original paper calls them annotations.} to 
(1) understand the degree to which commands can be parallelized (if at all) and 
(2) correctly transform shell pipelines into data-flow graphs using the command's input and output streams.

\heading{Parallelizability}
Determined by the way a command processes its input and output streams, a command can be classified into one of four parallelizability classes~\cite{dataflow:icfp2021}:
%
stateless, 
parallelizable pure, 
non-parallelizable pure, 
and side-effectful. 
Side-effectful commands mutate the filesystem state or depend on some state unrelated to their input (\eg \ttt{rm} and \ttt{pwd}).
The remaining three classes have no side effects and operate on well-defined input and output files, differing only in parallelizability.
Stateless commands process input line by line without maintaining state and pure commands preserve state across lines.
Optimization systems such as PaSh and POSH~\cite{dish, fractal:nsdi:2026} use these distinctions to determine when a command can be safely parallelized across input chunks.
Some pure commands can be parallelized using an aggregator and are therefore considered parallelizable pure (\eg \ttt{wc}) while others cannot (\eg \ttt{sha256sum}).
%
%
\sys can determine if a command is stateless, pure, or side-effectful, but cannot distinguish between non-parallelizable pure and pure commands, since that amounts to being able to synthesize an aggregator, which is out of scope for \sys but covered in prior work~\cite{nicolet-pldi17,nicolet-pldi19,nicolet-pldi21}.
%

\heading{Inputs and outputs}
A command invocation's input and output streams are derived from the its arguments or are specific file descriptors (\eg \ttt{stdin} and \ttt{stdout}).
For example, the \ttt{cat} invocation has \ttt{stdin} as input and \ttt{stdout} as output, while \ttt{grep "foo" f1} has the file \ttt{f1} as input and \ttt{stdout} as output.
Using the information from the system-call traces for each invocation, together with the input and output arguments, \sys is able to generate a specification of each command's inputs and outputs.
This information is used by shell optimization systems~\cite{pash_osdi,dish}, to determine inter-command data dependencies within shell pipelines.


\heading{POSH}
POSH~\cite{posh2020} is a system that accelerates I/O-heavy shell scripts by optimizing data movement.
Command specifications allow POSH to split a single command invocation to multiple distributed invocations across its input and arguments, and
minimize data movement.

\heading{Splittable input}
A single command invocation with $n$ arguments may be equivalent to concatenating $k$ invocations with $n / k$ arguments.
For instance, \ttt{cat}'s positional arguments are labeled \ttt{splittable} because \ttt{cat A B} is equivalent to \ttt{cat A} and \ttt{cat B}.
POSH uses this information to split an invocation into multiple ones that can be executed in parallel on different nodes, reducing data movement.

\heading{Input filtering}
Some commands are likely to produce outputs smaller than their inputs; for example, \ttt{grep} typically filters its input.
This property does not affect correctness but serves as a performance hint, allowing a system like POSH to avoid moving data across nodes before filtering occurs.

\heading{Location dependence}
Commands like \ttt{ls}, \ttt{git}, and \ttt{pwd} implicitly change their behavior based on the current working directory, irrespective of their arguments.
A system like POSH needs to be aware of this, in order to avoid offloading a command execution to a remote node where this implicit dependency is not satisfied.

\subsection{Bug Finding Systems}

This section describes two bug finding systems, Shellcheck and Shseer, and how \sys generates their specifications.

\heading{Shellcheck}
Shellcheck is a static analysis tool for shell scripts that finds common bugs and anti-patterns.

\heading{Command-specific checks}
Correctness checks that fall within scope are 
(1) checks related to incorrect usage of commands in terms of their arguments
and
(2) checks about negative command invocation effects, \eg invoking a command that deletes its arguments with a system directory. 
The first category can be captured by \sys's syntax specification, \ie if a command is invoked with types that are incompatible to the specification, and the second can be captured by determining a command's postconditions, \ie marking the arguments of a command that might be deleted by the command invocation.

\heading{Shseer}
Shseer~\cite{sash-hotos25} is a fully automated, semantics-driven, static-analysis system for finding bugs in shell scripts.
Its main focus is to detect bugs related to filesystem state, \ie cases where a script might accidentally delete a system directory.
In contrast to Shellcheck, Shseer does not perform a syntactic analysis but instead symbolically reasons about the execution of a script and its effects to the shell and filesystem state.
Given that all modifications to the filesystem state in a shell script happen through commands, Shseer critically depends on knowing a complete specification for each command invocation, \ie knowing exactly what its effects (postconditions) on the surrounding filesystem would be given some requirements on the starting filesystem (preconditions).

\heading{Pre- and post-conditions}
A command's behavior can be described in terms of its pre- and post-conditions.
For example, an invocation like \ttt{rm ARG} can only succeed if \ttt{ARG} points to a simple file in the filesystem; 
if \ttt{ARG} points to a nonexistent path or a directory, the command invocation will fail with a non-zero exit code.
Shseer uses annotations that describe these preconditions and postconditions to symbolically reason about the effects of commands on the filesystem state.
Command preconditions are the minimal requirements on argument types~(\cref{fig:fs-model}) that must hold for a command to run. 
Its postconditions then describe the actions the command performs on the filesystem and the command's exit status.
\sys derives these pre- and postconditions from the system-call traces and the filesystem state before and after command execution.




\subsection{Specification Derivation}
With execution traces in hand, \sys infers specifications using a set of rules and heuristics that are tailored to the target specifications.
This section describes the procedure with which \sys infers the specifications for each downstream system using the recorded system-call traces.


\heading{Parallelizability} 
\sys can assign to each command a parallelizability class $\mathcal{P}$: stateless, pure, or side-effectful.
Let C be a command invocation.
Let $C_i$ be a configuration of this invocation that reads from an input file $i$: either a regular file or \ttt{stdin}.
Let $\ttt{i}_1, \ldots \ttt{i}_n$ be $n$ partitions of $i$, split over lines.
If the command has side-effects (\ie writes to a path not named in its arguments), it is side-effectful.
If the command has no side-effects and if the output of invoking $C_i$ is identical to concatenating the outputs of invoking $C_{i_1}, \ldots, C_{i_n}$, the command is stateless.
Otherwise, it is pure.

\heading{Argument splittability}
Let $C$ be a command invocation with arguments $a_0, \ldots, a_n$ where $n \geq 2$, and let $C_i$ be an invocation identical to $C$ except only using $a_i$ as its argument.
If, after invoking $C_i$ for all $0 \leq i < n$, and concatenating their outputs, the result is identical to invoking $C$ with all arguments, then $C$ in terms of $a$ is considered splittable.

\heading{Input and output files}
Given a command invocation $C$ which has successfully executed under environment $E$ and $T$ is the set of traces extracted, if system call $r \in T$ , where $r_{name} \in R$ ($R$ being the set of read system calls), then $file(r)$ is considered an input file.
If system call $w \in T$ , where $w_{name} \in W$ ($W$ being the set of write system calls), $file(w)$ is considered an output file.
\sys produces sets $I$ and $O$ which denote all of the invocation's input and output files.

\heading{Output filtering}
Given a command invocation $C$, after \sys identifies sets $I$ and $O$ corresponding to the invocation's input and output files, $i_c$ is the content of an input file $i \in I$, $|i_c|$ is the size of $i_c$, and $o_c$ is the content of an output file $o \in O$, $|o_c|$ is the size of $o_c$.
If $\forall i \in I \; \forall o \in O:\; |o_c| < |i_c|$, the command invocation filters its input, and is marked as monotonically decreasing ($\downarrow$).

\heading{Current directory dependence}
Given a command invocation $C$
and the resulting set of traces $T$, if the system call $\text{\ttt{getcwd}} \in T$,
this command invocation's behavior is dependent on the current directory.

\heading{Path-oriented pre- and post-conditions}
Given a command execution $C$, an accompanying execution environment $E$, and the resulting set of traces $T$, \sys infers that the effects of running the command under environment $E$ will result in the set of traces $T$.
Specifically, \sys iterates over the predicates $E$ and sets their conjunction as the command pre-condition $P$.
\sys then constructs the command's post-condition $Q$ as the disjunction of all filesystem modifications in $T$ that were observed after its execution.






\subsection{Adapters}
\label{sec:adapters}
Partial command specifications generated by \sys 
are not immediately pluggable to downstream systems.
As a final step, \sys converts its intermediate specification format 
to each consumer system's required format. 
For PaSh and Shseer is a JSON-based language that encodes a command's properties 
related to its suitability to be converted to a DFG node and its filesystem side-effects, for POSH is a YAML-based language that encodes the relevant information, 
and for Shellcheck is a Haskell function that encodes a command's syntactic preconditions for a specific check.

\section{Evaluation}
\label{sec:eval}


To evaluate \sys's correctness and usefulness, we apply it on a suite of real-world commands to answer the following: 
\begin{enumerate}[label=\textbf{Q\arabic*}]
    \item What is the quality of specifications generated by \sys for each consumer system? (\cref{sec:eval:correctness})
    \item What is the accuracy of the syntax specifications generated by \sys's LLM-based parser? (\cref{sec:eval:llm-parser})
    \item How comprehensive is \sys's invocation generation in terms of real-world command usage? (\cref{sec:eval:completeness}) 
    \item How much time does \sys need to generate a specification for a command? (\cref{sec:eval:comp-cost})
\end{enumerate}

\heading{Implementation}
\sys is implemented in 6,520 lines of Python code, using OpenAI's GPT-4o~\cite{hurst2024gpt} model as the default LLM.
The system also uses \ttt{strace}~\cite{strace} to record system-call traces, and OverlayFS to provide an isolated filesystem environment for command execution.

\heading{Evaluation suite}
The first half of the suite consists of 88 specification ground truths from PaSh (52), POSH (17), Shellcheck (6), and Shseer (12), covering 60 unique commands (with overlap between the systems).
These commands span a broad range of domains such as text transformation, media encoding, and filesystem manipulation.
Among the original ground-truth specifications, the evaluation has excluded specifications for
(a) custom runtime-supporting commands without accompanying documentation (\eg PaSh's \ttt{eager} command) as \sys relies on documentation to explore a command's invocation space,
(b) shell builtins that do not have standalone binaries (\eg \ttt{history}, \ttt{read}), 
and 
(c) the \ttt{hdfs}~\cite{hdfs} command, as its filesystem model diverges significantly from \sys's.
Furthermore, we also apply \sys on an additional 60 commands from GNU coreutils for which we don't have access to ground-truth specifications to evaluate the robustness of the LLM Parser (Q3) and \sys's performance (Q4).


\subsection{Specification Correctness}
\label{sec:eval:correctness}

This section evaluates the correctness of specifications that \sys generates. 
We evaluate the specifications that \sys generates on two classes of systems: two performance optimization systems (PaSh and POSH) and two bug-finding systems (Shellcheck and Shseer).
Correctness is defined differently for each class of systems. 
For the performance optimization systems, divergence from the ground truth could lead to (1) incorrect results, \eg if a non-parallelizable command was specified as parallelizable, or (2) missing an optimization opportunity, \eg if a command was parallelizable but was not marked as such.
For the bug-finding systems divergence from the ground truth specification could lead to (1) false negatives, \ie missing a bug, or (2) false positives, \ie erroneously reporting something as a bug.
We use the existing command specifications offered by PaSh, POSH, Shellcheck, and Shseer to evaluate \sys's generated specifications.
\Cref{tab:correctness-summary} shows all the results.

\heading{Methodology: Performance optimization systems}
For the performance optimization systems we manually compare the command specifications generated by \sys to the ground truth ones that were handwritten for each system.
For PaSh, we also replace its built-in specifications with the ones generated by \sys and run its benchmark suite~\cite{pash_osdi} to check that the optimized scripts produce the same output as the original ones.
For POSH, as we were unable to run it, we directly compare the complete specification that \sys generates with the ones POSH comes with.
PaSh's specifications are more expressive than POSH since they describe command parallelizability for all possible flag and option invocations.
However, the specifications provided with the original system do not always generalize outside of their evaluation benchmark suite;
for example, \ttt{grep} is marked as always stateless in the original specifications, but in reality it is not if invoked with the \ttt{-c} flag (no invocation with \ttt{-c} exists in its benchmark suite).
Similar issues affect the ground-truth specifications provided alongside PaSh for the commands \ttt{ps} (marked as stateless, but in reality being side-effectful) and \ttt{cp} (marked as side-effectful, but in reality being pure).
%
Given this lack of generalization, the original PaSh specifications can only be considered ground-truths for the invocations that appear in its evaluation suite.
Thus, we compare \sys's generated specifications for those invocations only.

\heading{Results}
\sys successfully generates specifications for 68 out of 69 commands with ground truths in PaSh and POSH.
For PaSh, \sys correctly identifies a command's input and output streams (\eg \ttt{stdin}, \ttt{stdout}) and its parallelizability class.
Executing its benchmark suite with \sys's specifications also produces identical results (confirmed by hash comparison).
Note that as mentioned in \cref{sec:specs}, \sys does not synthesize command aggregators so we consider it correct if \sys can determine that a command is pure without requiring that it distinguishes between parallelizable and non-parallelizable pure.
For POSH, specifications have a boolean field that specifies if the command is splittable across its input stream.
This flag is semantically equivalent to a flag indicating whether the command invocation is stateless (as stateless is defined in PaSh's specifications) or not, and \sys successfully generates it.
\sys also captures the output of a command is shorter than its input, corresponding to POSH's \ttt{filters\_input} annotation.
For POSH, \sys is unable to generate any specifications for the \ttt{git} command, as its filesystem pre-conditions (\eg running inside a folder with a populated \ttt{.git} directory) is outside \sys's current filesystem model.
Finally, \sys covers thousands of command invocations absent from the built-in ground-truth specifications.
%


\heading{Methodology: Bug-finding Systems}
To evaluate \sys in the context of the bug-finding systems, we manually compare specifications against ground truths and test the systems using the generated specifications.
For manual comparison, we check equivalence of the generated specifications against the original ones (6 and 18 specifications for Shellcheck and Shseer respectively). 
Furthermore, to demonstrate their practical value, we check that running the systems on a set of tests with and without the specifications generated by \sys produces the same results.
%
Shellcheck originally includes 47 command-specific checks, out of which:
(1) 22 are related to subjective programming practices, \eg using the deprecated \ttt{egrep} command,  which \sys cannot determine without hints from the users;
(2) 10 are related to quoting and escaping issues, \eg not quoting the arguments of the \ttt{ssh} command, which are independent from the exact command syntax and specifications;
and (3) 9 are related to built-in commands, \eg the \ttt{exit} command accepts values in the 0--255 range, which cannot be analyzed by \sys.
The remaining 6 checks are in scope and \sys can generate specifications that can be used to directly generate the checks.
Since Shellcheck defines checks as Haskell functions, \sys first evaluates whether an invocation meets the check's conditions and the corresponding \sys adapter~(\cref{sec:adapters}) uses a check template which \sys then instantiates with the command name and arguments.
\sys also discovers new checks; \eg a catastrophic remove check for \ttt{rmdir}, and checks for commands that take two mandatory arguments like \ttt{chcon}, \ttt{comm} and \ttt{diff}.

For Shseer~\cite{sash-hotos25}, we execute the system using \sys-generated specifications on a set of 12 test scripts written by its authors and verify that Shseer correctly identifies the bug in each script.





\heading{Results}
For Shellcheck, \sys generates 6 of the supported command checks, which we manually insert into Shellcheck's source. 
The shellcheck test suite comprises over 2.2K tests (many of which are property tests, generating over 49K concrete checks in total), out of which 34 are related to the six checks that are generated by \sys.
While using the checks generated by \sys, Shellcheck produces the same output as when using the original checks, correctly passing all tests.
Lastly, \sys generates Shseer specifications by producing filesystem precondition and postcondition pairs, containing information the required filesystem state for a command to execute successfully (or not) and the effects of executing the command on the filesystem.
Shseer correctly reports all 12 bugs when used with the specifications generated by \sys.



\begin{table}[t]
\centering
\caption{\textbf{Correctness results.} Summary of correctness results of applying \sys to four state-of-the-art systems.
}
\begin{adjustbox}{max width=\columnwidth}
\begin{tabular}{llr}
\toprule
    System & Benchmark Suite & Correct \\
\midrule
    PaSh~\cite{pash_osdi} & PaSh hand-made specifications  & 52/52 (100\%)  \\
    POSH~\cite{posh2020} & POSH hand-made specifications & 16/17 (94\%) \\
    Shellcheck~\cite{holen_shellcheck_2024} & Shellcheck hand-made checks & 6/6 (100\%) \\
    Shseer~\cite{sash-hotos25} & Shseer hand-made specifications & 18/18 (100\%) \\
\bottomrule
\end{tabular}
\end{adjustbox}
\label{tab:correctness-summary}
\vspace{-1em}
\end{table}

\subsection{Syntax Specification Correctness}
\label{sec:eval:llm-parser}

To generate command syntax specifications \sys uses an LLM, which is not guaranteed to provide correct results.
In this section, we evaluate the correctness of the syntax specification generator across all 120 commands in the suite.


\heading{Ground-truth}
We created a ground-truth syntax specification for all 120 commands.
Two graduate students spent ~80 person-hours independently annotating \ttt{man} pages with argument types, then reconciled their results.
They found 16 discrepancies (4 arity errors, 12 overly general \ttt{String} types), which were resolved by re-examining the documentation, arriving at the final ground-truth set.


\heading{Methodology}
We compare the generated and ground-truth syntax specifications on an argument-by-argument basis.

\heading{Results}
Evaluating \sys's LLM parser component on 120 commands resulted in 116 perfect matches, with the rest of the four commands showing at least one discrepancy.
\sys's LLM component produced two classes of mistakes when translating documentation into syntax specifications:
type missclasifications (1 instance), and
missing or spurious options (3 instances).
Type missclasifications included typing the \ttt{-{}-delimiter} option in \ttt{cut} as \ttt{String} instead of \ttt{Char}.
Missing or spurious options include omitting an argument related to dereferencing command-line symlinks to directories for the \ttt{dir} command, missing three flags related batch size, compression program, and sorting key for the \ttt{sort} command, and missing a flag related to printing \ttt{pandoc}'s highlighting style for \ttt{pandoc}.


\begin{figure}[t]
    \centering
    \includegraphics[width=\columnwidth]{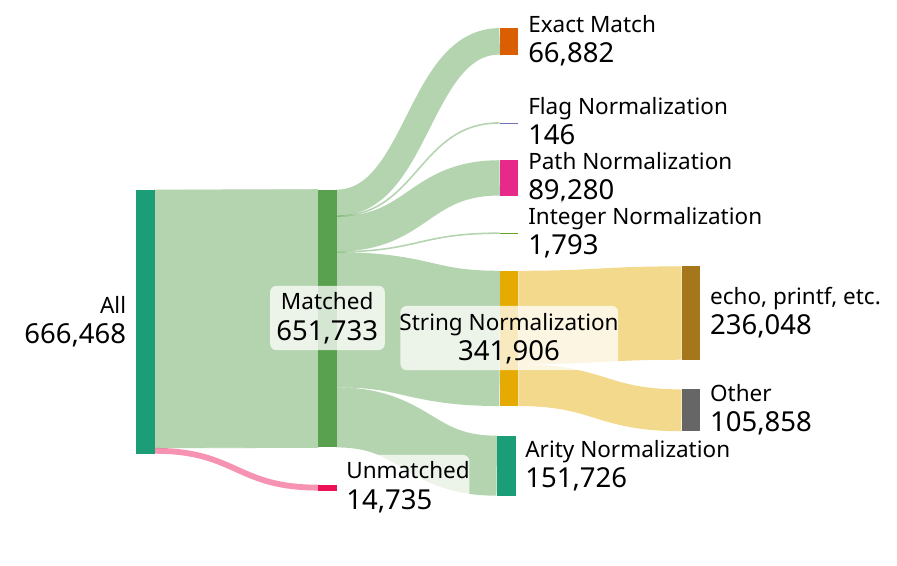}
    \caption{\textbf{Diagram of matched invocations.} Breakdown of the number of invocations found against the ones supported by \sys.
    }
    \label{fig:matches}
\end{figure}


\begin{figure*}[t]
    \includegraphics[width=\textwidth]{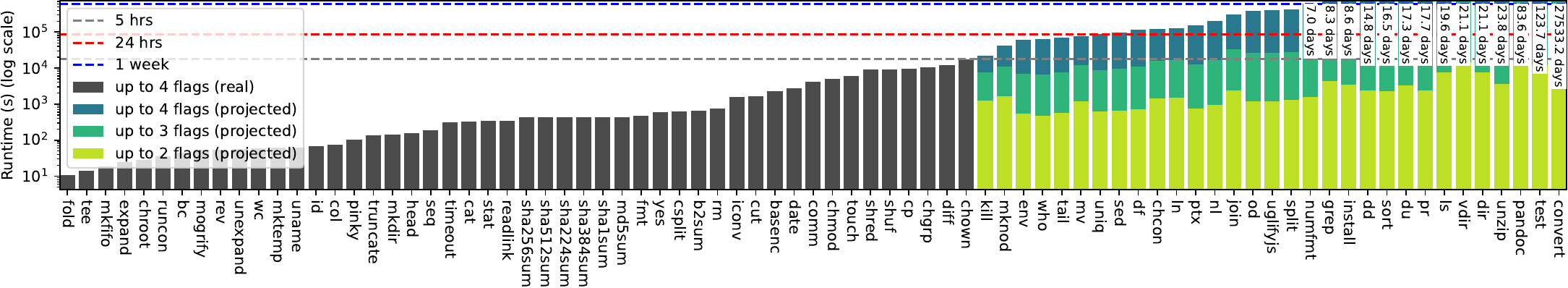}
    \caption{\textbf{Real and projected \sys execution time.}
    Bars in black are the real execution times of \sys with a maximum number of 4 flags or options. 
    Colored bars represent projected execution times with decreasing combination limits (up to 4, 3, and 2 flags).
    Omitted are 41 commands that took less than 10 seconds to apply \sys on.
    } 
    \label{fig:execution-times}
\end{figure*}

\subsection{Comprehensiveness}
\label{sec:eval:completeness}

To assess the comprehensiveness of \sys's invocation exploration and testing strategy, we use the set of command invocations that appear in the command invocation study~(\cref{sec:flag-study}). 
The distribution of the number of invocations per command is shown in \cref{fig:invdistr}. Command usage follows a power-law, since the top 10 commands amount to more than 85.95\% of all usages.

\heading{Methodology}
For each concrete invocation, we check and evaluate if \sys has tested against it during its invocation exploration.
Beyond literal syntactic equivalence, we compare invocations with some simple normalization rules, \eg flag normalization considers that \ttt{rm -rf} and \ttt{rm -fr} are equivalent, even though they are syntactically distinct.

\heading{Results}
\Cref{fig:matches} shows the number of exactly matched invocations, together with the invocations that match given a specific normalization.
Using all the normalization criteria, \sys matches 651,733 out of 666,468 invocations (97.78\%).
Out of all the 666,468 invocations, 66,882 (10\%) are exact matches---\sys executed the exact same invocation during its exploration.
Including flag normalization pushes this to a total of 64,542 matches.

The next normalization criteria refer to the types of arguments, and the generality of \sys's input generators~(\cref{sec:cmd-invocations}).
Path normalization describes that even though the invocations that \sys explored have different paths from the invocations found in the wild, they can still be considered equivalent since a command behavior depends not on a specific path name, but rather the identified location type (\eg file, directory, empty).
Similarly, integer normalization describes that even though \sys didn't explore all integer values for a command, it still accurately explored its behavior.
%
With path and integer normalization, the number of matched invocations goes to 122,816 (18\%).

String normalization refers to the rest of the arguments for which \sys only has a string type.
Including string normalization, \sys's matches increase to 632,780 (94\%) out of all invocations.
Importantly, this is not a valid normalization for commands whose semantics depend heavily on the input argument content (\eg higher order commands like \ttt{xargs} and \ttt{nohup}), \eg xargs will execute an arbitrary command depending on its string arguments, so running it with a subset of all strings doesn't adequately cover its behavior.
%
However, many of the commands in the dataset only use string arguments for printing and filtering, \eg \ttt{echo}, \ttt{printf}, \ttt{grep}, \ttt{tr}.
Including just these command invocations, \sys still matches 236,048 instances. 

Finally, commands have arguments with variable arity, \eg \ttt{rm} can be applied to an arbitrary number of pathnames.
Arity normalization assumes that a command does not change its behavior if given different numbers of variable arity arguments;
note that \sys is still able to extrapolate that  \ttt{rm f1 f2 f3 f4} will delete \ttt{f1}, \ttt{f2}, \ttt{f3}, and \ttt{f4} even if it has not invoked it with 4 arguments.
However, if a command changes behavior depending on arity, \eg if \ttt{rm f1 f2 f3 f4 f5} also wrote to a file called \ttt{deleted-5-files.txt}, then \sys wouldn't be able to determine that.
With all normalizations applied, \sys covers \textbf{651,733, or 97.78\%} of all collected command invocations.

\subsection{Computational Cost}
\label{sec:eval:comp-cost}

This section evaluates the time required for \sys to generate complete specifications.
We use the findings of the command invocation study~(\cref{sec:flag-study}) to only explore command invocations using up to four flags, since this captures more than 99.998\% of all invocations in our GitHub study.
We ran \sys on all 120 commands with a timeout of five hours per command. 
For the commands that didn't finish in the five hour mark, we extrapolate how long they would take to run for up to two, three, and four flags and options.
To give a sense of an upper bound, we use the 95th percentile of the time \sys takes per configuration for the projections based on the partial run, and report the number of configurations \sys would need to explore for each command times this time.
The experiments were run on an Intel Xeon E5-2667 v2 running Ubuntu 22.07 with Python 3.11.
%
%

\heading{Results}
%
%
\Cref{fig:execution-times} shows the results: with combination limit of four flags, \sys successfully generates specifications for 80 commands within an hour and 96 within a day, while 24 take more than a day.
Limiting the combination limit to two flags, which still captures over 99\% of all invocations in the study~(\cref{sec:flag-study}), 103 commands finish within an hour, and all but one commands finish within a day.
The sole outlier is \ttt{convert}, with a projected runtime of 98.72 hours for two flags.
This result is not particularly concerning: \ttt{convert} is a command that has a simple specification,
\ie it just reads a single file which it converts and writes to a different output file,
but has a very complex interface, \ie many flags and options that slightly modify the conversion that happens without affecting its specification.
Since its semantics remain essentially the same, the exhaustive flag combinations exaggerate its complexity without revealing meaningful behavioral variation.

\sys's runtime can be further reduced, as it can be trivially parallelized, as each configuration exploration is independent of the others.
Currently, \sys uses all available CPU cores to parallelize configuration tracing within a single machine.
%
\section{Related Work}
\label{sec:related-work}

\heading{Test generation \& fuzzing}
Test generation and fuzzing systems \cite{KLEE_2008, godefroid2012sage, gupta2022clifuzzer, afl, libfuzzer} use various techniques to generate test cases 
with the end goal of identifying program misbehavior.
Notably, KLEE~\cite{KLEE_2008} acts on LLVM bytecode and uses concolic execution to explore most possible execution paths,
attempting to generate inputs that will reach each path
and has shown success in testing projects like GNU coreutils.
\sys differs from these systems as it does not make assumptions about the language a component is written in, but rather leverages documentation to guide generation of typical-usage tests.
That said, this insight is related to black-box test generation systems that leverage design documentation and specifications~\cite{tvkb-compasc-2001} or keywords~\cite{dac-icqs-2014} to generate tests without source-code knowledge, and could inform initial test generation strategies even for systems that do leverage source/coverage knowledge.
\sys also differs in intent: it does not attempt to identify edge cases or test correctness, but rather
treats commands as correct and generates specifications based on observed behavior.


\heading{Specification mining}
Specification mining~\cite{ammons2002mining, lemieux2015general, li2010scalable, lee2011mining, shoham2007static} is the process of converting traces to higher-level 
representations that identify temporal and functional properties of a given program in an abstract form.
\sys differs from prior work in two key ways.
First, it traces observable events produced opaque, unmodified command binaries rather than instrumenting conventional APIs or code, and targets properties that matter for systems-level reasoning (\eg side effects, dependencies, monotonicity, \etc).
Second, unlike most earlier systems that rely on user-provided inputs or execution traces, \sys automatically generates command invocations from documentation, allowing it to explore and mine specifications without requiring test cases or nominal executions.
Hence, \sys's approach for generating execution traces could be useful for other specification settings and systems more broadly.

\heading{Invariant inference}
Invariant inference systems~\cite{ernst2007daikon, dysy-icse-2008} use static and/or dynamic analysis to analyze source code and extract program invariants on a program statement level.
These systems often convert source code into an intermediate form which they then analyze to extract invariants.
\sys differs from these systems as it is language agnostic, using documentation to identify valid invocations and operating-system-level tracing to extract behavior traces, and operates at a command-invocation granularity.
That said, \sys's approach of extracting unknown information from documentation could be leveraged in the contexts of invariant inference systems to extract hypothesis invariants to be analyzed/checked.

\heading{Program analysis}
Program analysis techniques, such as predicate mining
\cite{ramanathan2007static} and automata-based abstractions
\cite{shoham2007static}, have been employed to extract formal specifications
from source code. 
These methods typically operate on statically analyzable
programs written in specific languages (\eg Java) and infer behavioral
properties like valid API usage patterns or invariants.
In contrast, \sys adopts a language-agnostic approach, treating commands as
black boxes without relying on their implementation details.
\sys
extracts high-level, human-interpretable specifications that capture
functional properties of commands. 

\heading{\unix Synthesis}
Prior work on synthesis for \unix shell commands and pipelines~\cite{macho:13,bhansali1993synthesis} extracts partial specifications from natural-language documentation.
Instead of automatically generating parallel or distributed versions of an
existing command or pipeline, \sys's goal is to infer command specifications for 
all of a command's flags and options, usable by downstream systems.

\heading{Documentation mining}
Many prior attempts at automatically extracting structured information from natural language have used manually-written rules and heuristics.
The most widely used example is Explainshell~\cite{explainshell}, a popular open source library which dissects a given command invocation and maps its parts into flags and options from its \ttt{man} page using a combination of string processing and a classification model.
This approach, however, is limited cannot always track and reason about nuances in a command's syntax and the type of its arguments.
%
\sys instead applies an LLM to transform the documentation into a structured representation
which encodes the syntactically and semantically valid ways to invoke a command.
\section{Conclusion}
\label{sec:conclusion}

This paper presents \sys, a fully automatic system for generating partial specifications of opaque software components and focuses on its application to shell commands.
%
\sys's evaluation demonstrates that it is possible to automatically generate reusable and accurate specifications for opaque commands, addressing a key bottleneck in systems requiring extensive annotations.
%

\begin{acks}
We'd like to thank Anirudh Narsipur, Julian Dai, and Grigoris Ntousakis for early discussions on some of the ideas in this paper.
Material is based upon research supported by NSF awards CNS-2247687 and CNS-2312346,
  DARPA contract no. HR001124C0486,
  an Amazon Research Award (Fall 2024),
  a Google ML-and-Systems Junior Faculty Award,
  a seed grant from Brown University's Data Science Institute,
  and a BrownCS Faculty Innovation Award.
\end{acks}

\bibliographystyle{plain}
\bibliography{bib}

\begin{thebibliography}{10}

\bibitem{ammons2002mining}
Glenn Ammons, Rastislav Bod{\'\i}k, and James~R Larus.
\newblock Mining specifications.
\newblock {\em ACM Sigplan Notices}, 37(1):4--16, 2002.

\bibitem{becker:hal-03737886}
Benedikt Becker, Nicolas Jeannerod, Claude March{\'e}, Yann R{\'e}gis-Gianas,
  Mihaela Sighireanu, and Ralf Treinen.
\newblock {The CoLiS Platform for the Analysis of Maintainer Scripts in Debian
  Software Packages}.
\newblock {\em {International Journal on Software Tools for Technology
  Transfer}}, 2022.

\bibitem{bhansali1993synthesis}
Sanjay Bhansali and Mehdi~T Harandi.
\newblock Synthesis of unix programs using derivational analogy.
\newblock {\em Machine Learning}, 10(1):7--55, 1993.

\bibitem{nosilverbullet}
Brooks.
\newblock No silver bullet essence and accidents of software engineering.
\newblock {\em Computer}, 20(4):10--19, 1987.

\bibitem{KLEE_2008}
Cristian Cadar, Daniel Dunbar, and Dawson Engler.
\newblock Klee: unassisted and automatic generation of high-coverage tests for
  complex systems programs.
\newblock In {\em Proceedings of the 8th USENIX Conference on Operating Systems
  Design and Implementation}, OSDI'08, page 209–224, USA, 2008. USENIX
  Association.

\bibitem{macho:13}
Anthony Cozzie, Murph Finnicum, and Samuel~T King.
\newblock Macho: Programming with man pages.
\newblock In {\em 13th Workshop on Hot Topics in Operating Systems}, Napa, CA,
  United States, May 2011. USENIX Association.

\bibitem{dysy-icse-2008}
Christoph Csallner, Nikolai Tillmann, and Yannis Smaragdakis.
\newblock Dysy: dynamic symbolic execution for invariant inference.
\newblock In {\em Proceedings of the 30th International Conference on Software
  Engineering}, ICSE '08, page 281–290, New York, NY, USA, 2008. Association
  for Computing Machinery.

\bibitem{dac-icqs-2014}
Mohammad~Ali Darvish~Darab and Carl~K. Chang.
\newblock Black-box test data generation for gui testing.
\newblock In {\em 2014 14th International Conference on Quality Software},
  pages 133--138, 2014.

\bibitem{json-generator}
Microsoft~Edge DevRel.
\newblock Json dummy data generator, 2024.

\bibitem{ubuntuoxidising}
Ubuntu Discourse.
\newblock Carefully but purposefully oxidising ubuntu, 2024.
\newblock Accessed: 2024-12-10.

\bibitem{linuxnamespaces}
Linux~Kernel Documentation.
\newblock Linux namespaces, 2024.
\newblock Available at
  \url{https://man7.org/linux/man-pages/man7/namespaces.7.html}.

\bibitem{overlayfslinux}
Linux~Kernel Documentation.
\newblock Overlayfs - linux kernel documentation, 2024.
\newblock Available at
  \url{https://www.kernel.org/doc/html/latest/filesystems/overlayfs.html}.

\bibitem{seccomp}
Linux~Kernel Documentation.
\newblock Seccomp - linux kernel documentation, 2024.
\newblock Available at
  \url{https://docs.kernel.org/userspace-api/seccomp_filter.html}.

\bibitem{ernst2007daikon}
Michael~D Ernst, Jeff~H Perkins, Philip~J Guo, Stephen McCamant, Carlos
  Pacheco, Matthew~S Tschantz, and Chen Xiao.
\newblock The daikon system for dynamic detection of likely invariants.
\newblock {\em Science of computer programming}, 69(1-3):35--45, 2007.

\bibitem{nicolet-pldi17}
Azadeh Farzan and Victor Nicolet.
\newblock Synthesis of divide and conquer parallelism for loops.
\newblock In {\em Proceedings of the 38th ACM SIGPLAN Conference on Programming
  Language Design and Implementation}, PLDI 2017, page 540–555, New York, NY,
  USA, 2017. Association for Computing Machinery.

\bibitem{nicolet-pldi19}
Azadeh Farzan and Victor Nicolet.
\newblock Modular divide-and-conquer parallelization of nested loops.
\newblock In {\em Proceedings of the 40th ACM SIGPLAN Conference on Programming
  Language Design and Implementation}, PLDI 2019, page 610–624, New York, NY,
  USA, 2019. Association for Computing Machinery.

\bibitem{nicolet-pldi21}
Azadeh Farzan and Victor Nicolet.
\newblock Phased synthesis of divide and conquer programs.
\newblock In {\em Proceedings of the 42nd ACM SIGPLAN International Conference
  on Programming Language Design and Implementation}, PLDI 2021, page
  974–986, New York, NY, USA, 2021. Association for Computing Machinery.

\bibitem{strace}
Paul Floyd et~al.
\newblock strace - the linux syscall tracer, 2023.
\newblock Accessed: 2024-06-20.

\bibitem{godefroid2012sage}
Patrice Godefroid, Michael~Y Levin, and David Molnar.
\newblock Sage: Whitebox fuzzing for security testing: Sage has had a
  remarkable impact at microsoft.
\newblock {\em Queue}, 10(1):20--27, 2012.

\bibitem{gupta2022clifuzzer}
Abhilash Gupta, Rahul Gopinath, and Andreas Zeller.
\newblock Clifuzzer: mining grammars for command-line invocations.
\newblock In {\em Proceedings of the 30th ACM Joint European Software
  Engineering Conference and Symposium on the Foundations of Software
  Engineering}, ESEC/FSE 2022, page 1667–1671, New York, NY, USA, 2022.
  Association for Computing Machinery.

\bibitem{hdfs}
{Hadoop Project}.
\newblock {\em Hadoop Distributed File System}.
\newblock {Apache Software Foundation}, 2024.
\newblock Accessed: 2024-12-07.

\bibitem{dataflow:icfp2021}
Shivam Handa, Konstantinos Kallas, Nikos Vasilakis, and Martin~C. Rinard.
\newblock An order-aware dataflow model for parallel unix pipelines.
\newblock {\em Proc. ACM Program. Lang.}, 5(ICFP), August 2021.

\bibitem{holen_shellcheck_2024}
Vidar Holen and contributors.
\newblock Shellcheck, a static analysis tool for shell scripts, 2024.
\newblock Accessed: 2024-06-19.

\bibitem{fractal:nsdi:2026}
Zhicheng Huang, Ramiz Dundar, Yizheng Xie, Konstantinos Kallas, and Nikos
  Vasilakis.
\newblock Fractal: Fault-tolerant shell-script distribution.
\newblock In {\em 23rd USENIX Symposium on Networked Systems Design and
  Implementation (NSDI 26)}, Renton, WA, May 2026. USENIX Association.

\bibitem{hurst2024gpt}
Aaron Hurst, Adam Lerer, Adam~P Goucher, Adam Perelman, Aditya Ramesh, Aidan
  Clark, AJ~Ostrow, Akila Welihinda, Alan Hayes, Alec Radford, et~al.
\newblock Gpt-4o system card.
\newblock {\em arXiv preprint arXiv:2410.21276}, 2024.

\bibitem{pash_osdi}
Konstantinos Kallas, Tammam Mustafa, Jan Bielak, Dimitris Karnikis,
  Thurston~H.Y. Dang, Michael Greenberg, and Nikos Vasilakis.
\newblock Practically correct, {Just-in-Time} shell script parallelization.
\newblock In {\em 16th USENIX Symposium on Operating Systems Design and
  Implementation (OSDI 22)}, pages 769--785, Carlsbad, CA, July 2022. USENIX
  Association.

\bibitem{explainshell}
Idan Kamara and contributors.
\newblock explainshell.com - match command-line arguments to their help text,
  2024.
\newblock Accessed: 2024-10-25.

\bibitem{lampson2004software}
Butler Lampson.
\newblock {\em Software Components: Only The Giants Survive}, pages 137--146.
\newblock Springer Verlag, January 2004.
\newblock This paper was written for a symposium in honor of Roger Needham,
  February 2003. It is based on a keynote address at the 21st International
  Conference on Software Engineering, Los Angeles, California, 16-22 May 1999.

\bibitem{sash-hotos25}
Lukas Lazarek, Seong-Heon Jung, Evangelos Lamprou, Zekai Li, Anirudh Narsipur,
  Eric Zhao, Michael Greenberg, Konstantinos Kallas, Konstantinos Mamouras, and
  Nikos Vasilakis.
\newblock From ahead-of- to just-in-time and back again: Static analysis for
  unix shell programs.
\newblock In {\em Proceedings of the 2025 Workshop on Hot Topics in Operating
  Systems}, HotOS '25, page 88–95, New York, NY, USA, 2025. Association for
  Computing Machinery.

\bibitem{lee2011mining}
Choonghwan Lee, Feng Chen, and Grigore Ro\c{s}u.
\newblock Mining parametric specifications.
\newblock In {\em Proceedings of the 33rd International Conference on Software
  Engineering}, ICSE '11, page 591–600, New York, NY, USA, 2011. Association
  for Computing Machinery.

\bibitem{lemieux2015general}
Caroline Lemieux, Dennis Park, and Ivan Beschastnikh.
\newblock General ltl specification mining (t).
\newblock In {\em 2015 30th IEEE/ACM International Conference on Automated
  Software Engineering (ASE)}, pages 81--92. IEEE, 2015.

\bibitem{li2010scalable}
Wenchao Li, Alessandro Forin, and Sanjit~A Seshia.
\newblock Scalable specification mining for verification and diagnosis.
\newblock In {\em Proceedings of the 47th design automation conference}, pages
  755--760, 2010.

\bibitem{libfuzzer}
{LLVM Project}.
\newblock {libFuzzer – a library for coverage-guided fuzz testing}.
\newblock Web page, 2018.
\newblock Accessed 2025-09-25.

\bibitem{abash}
Karl Mazurak and Steve Zdancewic.
\newblock Abash: finding bugs in bash scripts.
\newblock In {\em Proceedings of the 2007 Workshop on Programming Languages and
  Analysis for Security}, PLAS '07, page 105–114, New York, NY, USA, 2007.
  Association for Computing Machinery.

\bibitem{dish}
Tammam Mustafa, Konstantinos Kallas, Pratyush Das, and Nikos Vasilakis.
\newblock {DiSh}: Dynamic {Shell-Script} distribution.
\newblock In {\em 20th USENIX Symposium on Networked Systems Design and
  Implementation (NSDI 23)}, pages 341--356, Boston, MA, April 2023. USENIX
  Association.

\bibitem{shash}
Anirudh Narsipur.
\newblock Towards automated reasoning for shell programs.
\newblock Technical report, Brown University, 2024.

\bibitem{unixenv}
R.~Pike and B.~W. Kernighan.
\newblock The unix system: Program design in the unix environment.
\newblock {\em AT\&T Bell Laboratories Technical Journal}, 63(8):1595--1605,
  1984.

\bibitem{posh2020}
Deepti Raghavan, Sadjad Fouladi, Philip Levis, and Matei Zaharia.
\newblock {POSH}: A {Data-Aware} shell.
\newblock In {\em 2020 USENIX Annual Technical Conference (USENIX ATC 20)},
  pages 617--631. USENIX Association, July 2020.

\bibitem{ramanathan2007static}
Murali~Krishna Ramanathan, Ananth Grama, and Suresh Jagannathan.
\newblock Static specification inference using predicate mining.
\newblock {\em ACM SIGPLAN Notices}, 42(6):123--134, 2007.

\bibitem{shoham2007static}
Sharon Shoham, Eran Yahav, Stephen Fink, and Marco Pistoia.
\newblock Static specification mining using automata-based abstractions.
\newblock In {\em Proceedings of the 2007 International Symposium on Software
  Testing and Analysis}, pages 174--184, 2007.

\bibitem{htmx}
Big~Sky Software.
\newblock Htmx, 10 2024.

\bibitem{tvkb-compasc-2001}
L.H. Tahat, B.~Vaysburg, B.~Korel, and A.J. Bader.
\newblock Requirement-based automated black-box test generation.
\newblock In {\em 25th Annual International Computer Software and Applications
  Conference (COMPSAC) 2001}, pages 489--495, 2001.

\bibitem{pash2021eurosys}
Nikos Vasilakis, Konstantinos Kallas, Konstantinos Mamouras, Achilles
  Benetopoulos, and Lazar Cvetkovi\'{c}.
\newblock Pash: light-touch data-parallel shell processing.
\newblock In {\em Proceedings of the Sixteenth European Conference on Computer
  Systems}, EuroSys '21, page 49–66, New York, NY, USA, 2021. Association for
  Computing Machinery.

\bibitem{afl}
Michal Zalewski.
\newblock American fuzzy lop (afl).
\newblock Whitepaper, 2013.
\newblock Accessed 2025-09-25.

\end{thebibliography}





\end{document}